\documentclass{ar2e}
\usepackage{txfonts}    
\usepackage{graphicx}
\usepackage{slashbox}
\usepackage{rotating}
\usepackage[usenames]{color}

\begin{document}
\newcommand{\definition}{{\bf \textcolor{BrickRed} {\bf Definition:}\,}}

\input epsf.tex
\input epsf.def

\jname{Ann. Rev. Astronomy and Astrophysics}
\jyear{2012}
\jvol{50}
\ARinfo{1056-8700/97/0610-00}

\title{Pre-Supernova Evolution of Massive Single and Binary Stars}

\markboth{Massive Stars}{Massive Stars}

\author{N. Langer
\affiliation{Argelander-Institut f\"ur Astronomie, Universit\"at Bonn, Germany}}


\begin{abstract}
Massive stars are essential to understand a variety of branches of astronomy including
galaxy and star cluster
evolution, nucleosynthesis and supernovae, pulsars and black holes. It has become evident that massive
star evolution is very diverse, being sensitive to metallicity, binarity, rotation, and possibly
magnetic fields. While the problem to obtain a good statistical observational database is alleviated by
current large spectroscopic surveys, it remains a challenge to model these diverse paths of massive stars
towards their violent end stage.

We show that the main sequence stage offers the best opportunity to gauge the relevance of the various
possible evolutionary scenarios. This also allows to sketch the post-main sequence evolution of massive
stars, for which observations of Wolf-Rayet stars give essential clues.
Recent supernova discoveries due to the current
boost in transient searches allow tentative mappings of progenitor models with supernova types,
including pair instability supernovae and gamma-ray bursts.
\end{abstract}

\maketitle
%

Find the full article at Annual Reviews of Astronomy and Astrophysics, Vol 50, pp. 107-164; 
2012, or directly at \\
{\tt http://www.annualreviews.org/doi/abs/10.1146/annurev-astro-081811-125534}

\end{document}